%% file: neutrino_ml.tex
\documentclass[a4paper,11pt]{article}
\pdfoutput=1 % if your are submitting a pdflatex (i.e. if you have
\usepackage{jinstpub} % for details on the use of the package, please
\usepackage{lineno}
\usepackage{comment}
\usepackage[bottom]{footmisc} 
\usepackage{footnote}

\newcommand{\el}[2]{\ensuremath{^{#2}\textrm{#1}}}

\newcommand{\nuebar}{\ensuremath{\overline{\nu}_{e}}}

\newcommand{\mevee}{\ensuremath{\textrm{MeV}}}
\newcommand{\pspt}{PROSPECT}

\title{\boldmath Machine Learning for Single-Ended Event Reconstruction in PROSPECT Experiment}

\affiliation[1]{Department of Physics, Boston University, Boston, MA, USA} \affiliation[2]{Brookhaven National Laboratory, Upton, NY, USA} \affiliation[3]{Department of Physics, Drexel University, Philadelphia, PA, USA} \affiliation[4]{George W.,Woodruff School of Mechanical Engineering, Georgia Institute of Technology, Atlanta, GA, USA} \affiliation[5]{Department of Physics and Astronomy, University of Hawaii, Honolulu, HI, USA} \affiliation[6]{Department of Physics, Illinois Institute of Technology, Chicago, IL, US} 
\affiliation[7]{Nuclear and Chemical Sciences Division, Lawrence Livermore National Laboratory, Livermore, CA, USA} \affiliation[8]{Department of Physics, Le Moyne College, Syracuse, NY, USA} \affiliation[9]{National Institute of Standards and Technology, Gaithersburg, MD, USA} \affiliation[10]{High Flux Isotope Reactor, Oak Ridge National Laboratory, Oak Ridge, TN, USA} \affiliation[11]{Physics Division, Oak Ridge National Laboratory, Oak Ridge, TN, USA} \affiliation[12]{Department of Physics, Susquehanna University, Selinsgrove, PA, USA} \affiliation[13]{Department of Physics and Astronomy, University of Tennessee, Knoxville, TN, USA} \affiliation[14]{Department of Physics, United States Naval Academy, Annapolis, MD, USA} \affiliation[15]{Department of Physics, University of Wisconsin, Madison, WI, USA} \affiliation[16]{Wright Laboratory, Department of Physics, Yale University, New Haven, CT, USA}

\author[6]{M.~Andriamirado,} \author[15]{A.~B.~Balantekin,} \author[8]{C.~D.~Bass,} \author[6]{O. Benevides Rodrigues,} \author[7]{E.~P.~Bernard,} \author[7]{N.~S.~Bowden,} \author[10]{C.~D.~Bryan,} \author[14]{R.~Carr,} \author[7]{T.~Classen,} \author[10]{A.~J.~Conant,} \author[10]{G.~Deichert,}
\author[11]{A.~Delgado,}
\author[3]{M.~J.~Dolinski,} \author[4]{A.~Erickson,} \author[10]{M.~Fuller,} \author[11,13]{A.~Galindo-Uribarri,} \author[2]{S.~Gokhale,} \author[1]{C.~Grant,} \author[2]{S.~Hans,} \author[12]{A.~B.~Hansell,} \author[9]{T.~E.~Haugen,} \author[16]{K.~M.~Heeger,} \author[11,13]{B.~Heffron,} \author[2]{D.~E.~Jaffe,} \author[3]{S.~Jayakumar,} \author[5]{J.~Koblanski,} \author[1]{P.~Kunkle,} \author[3]{C.~E.~Lane,} \author[6]{B.~R.~Littlejohn,} \author[3]{A.~Lozano Sanchez,} \author[11,13]{X.~Lu,} \author[6]{F.~Machado,} \author[5]{J.~Maricic,} \author[7]{M.~P.~Mendenhall,} \author[5]{A.~M.~Meyer,} \author[5]{R.~Milincic,} \author[11]{P.~E.~Mueller,} \author[9]{H.~P.~Mumm,} \author[3]{R.~Neilson,} \author[7]{C.~Roca,} \author[2]{R.~Rosero,} \author[11,13]{D.~Venegas-Vargas,} \author[16]{J.~Wilhelmi,} \author[2]{M.~Yeh,} \author[2]{C.~Zhang,} \author[7]{and X.~Zhang}

\emailAdd{baheffron@gmail.com}

\abstract{The Precision Reactor Oscillation and Spectrum Experiment, PROSPECT, was a segmented antineutrino detector that successfully operated at the High Flux Isotope Reactor in Oak Ridge, TN, during its 2018 run. 
Despite challenges with photomultiplier tube base failures affecting some segments, innovative machine learning approaches were employed to perform position and energy reconstruction, and particle classification. This work highlights the effectiveness of convolutional neural networks and graph convolutional networks in enhancing data analysis. By leveraging these techniques, a 3.3\% increase in effective statistics was achieved compared to traditional methods, showcasing their potential to improve analysis performance. Furthermore, these machine learning methodologies offer promising applications for other segmented particle detectors, underscoring their versatility and impact. }


\arxivnumber{2503.06727} % only if you have one

\collaboration[c]{on behalf of the PROSPECT collaboration}

\begin{document}
\maketitle
% Add the footnote to the first page 

\flushbottom 
\input{paper_jinst_revisions.tex}

\section*{Acknowledgement}
This material is based upon work supported by the following sources: US Department of Energy (DOE) Office of
Science, Office of High Energy Physics under Award No.
DE-SC0016357 and DE-SC0017660 to Yale University, under Award No. DE-SC0017815 to Drexel University, under
Award No. DE-SC0008347 to Illinois Institute of Technology,
under Award No. DE-SC0010504 to University of Hawaii,
under Contract No. DE-SC0012704 to Brookhaven National
Laboratory, and under Work Proposal Number SCW1504 to
Lawrence Livermore National Laboratory. This work was performed under the auspices of the U.S. Department of Energy
by Lawrence Livermore National Laboratory under Contract
DE-AC52-07NA27344 and by Oak Ridge National Laboratory under Contract DE-AC05-00OR22725. Additional funding for the experiment was provided by the Heising-Simons Foundation under Award No. \#2016-117 to Yale University.
We further acknowledge support from Yale University, the
Illinois Institute of Technology, Temple University, University
of Hawaii, Brookhaven National Laboratory, the Lawrence
Livermore National Laboratory LDRD program, the National
Institute of Standards and Technology, and Oak Ridge National Laboratory. We gratefully acknowledge the support and
hospitality of the High Flux Isotope Reactor and Oak Ridge
National Laboratory, managed by UT-Battelle for the U.S. Department of Energy.

The United States Government retains and the publisher, by accepting the article for publication, acknowledges that the United States Government retains a non-exclusive, paid-up, irrevocable, world-wide license to publish or reproduce the published form of this manuscript, or allow others to do so, for the United States Government purposes. The Department of Energy will provide public access to these results of federally sponsored research in accordance with the DOE Public Access Plan. This research used resources at the High Flux Isotope Reactor, a DOE Office of Science User Facility operated by Oak Ridge National Laboratory. This work was partially supported by the Department of Energy Office of High Energy Physics under FWP ERKAP60.

\bibliography{ml} 
\bibliographystyle{JHEP}
\end{document}

%% file: paper_jinst_revisions.tex
\section{Introduction}

The \pspt~experiment measured the \nuebar~emitted by the High Flux Isotope Reactor in 2018 to measure possible oscillation effects due to sterile neutrinos~\cite{PROSPECT:2018dtt,PROSPECT:2018snc,PROSPECT:2020sxr}.
Reactor \nuebar~were detected through the inverse beta-decay reaction (IBD): \nuebar + p → e+ + n, where the incoming antineutrino interacts with a proton to produce a positron and neutron.
The outgoing positron, carrying most of the initial \nuebar~energy (typically 2-8 MeV), provides a prompt signal.
The neutron, after thermalization, is captured several microseconds later providing a delayed signal.
This characteristic double-coincidence signature helps distinguish IBD events from numerous backgrounds, including cosmic rays, natural radioactivity, and reactor-induced events.

\begin{figure}[htbp]
	\centering
    \includegraphics[clip=true, trim=0mm 50mm 0mm 40mm,width=0.99\textwidth]{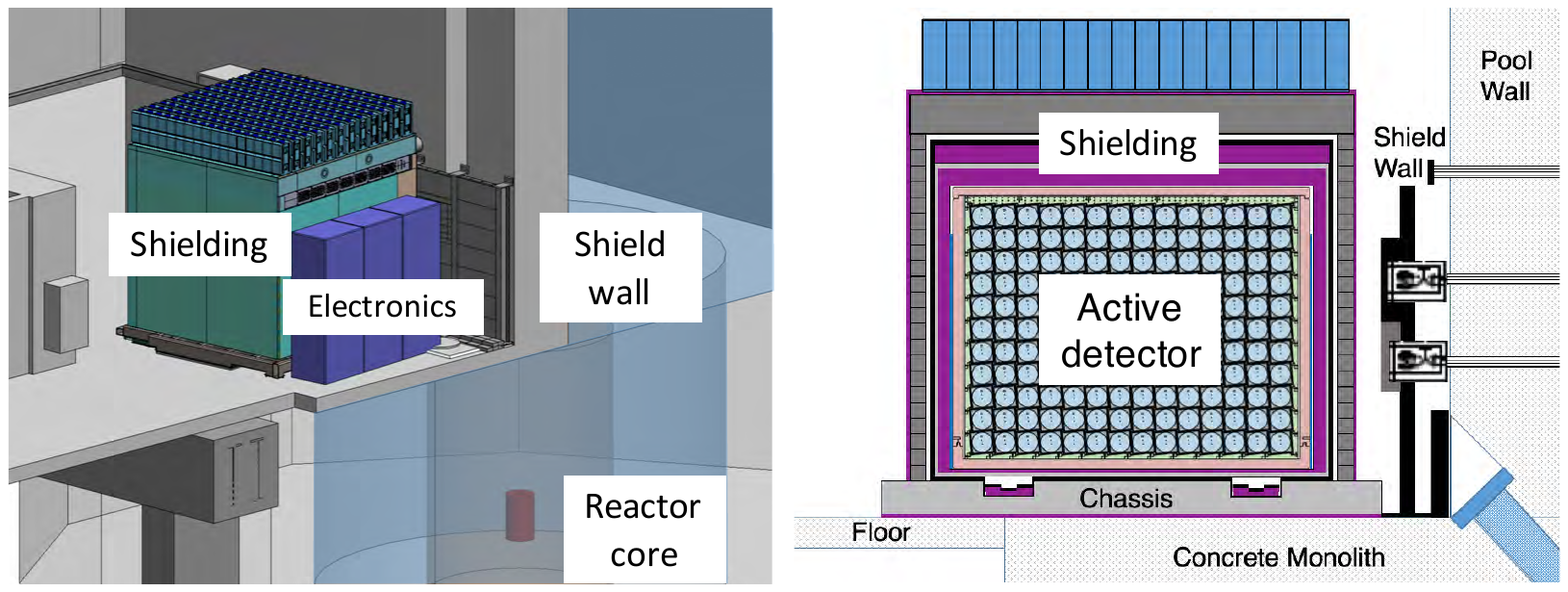}
	\caption{(left) Layout of the PROSPECT experiment. The detector is installed in the HFIR Experiment Room next to the water pool and 5 m above the HFIR
		reactor core (red). The floor below contains multiple neutron beam-lines and scattering experiments. (Right) Schematic showing the active detector volume divided
		into 14 (long) by 11 (tall) separate segments and surrounded by nested containment vessels and shielding layers. Shield walls cover penetrations in the pool wall
		associated with high backgrounds.~\cite{PROSPECT2018}
	}
	\label{fig:Detector}
\end{figure}

Measuring possible \nuebar~oscillations requires sufficient position resolution and suppression of backgrounds.
PROSPECT accomplished this with a detector consisting of an 11$\times$14 array of optically separated liquid scintillator segments outfitted with PMTs at either end of the segment, allowing cm-scale precision in the localization of energy deposits~\cite{prospect_optical_grid}.
See Figure~\ref{fig:Detector} for a schematic of the detector along with its position relative to the reactor.
The position along the length of the segment (henceforth $z$) was estimated utilizing the timing difference and relative amplitude of the light detected by the two PMTs.

To enhance neutron detection efficiency and background discrimination, the liquid scintillator was doped with \el{Li}{6}.
This isotope has a high thermal neutron capture cross section, which allows neutron identification through the \el{Li}{6}(n,$\alpha$)\el{H}{3} reaction with a Q-value of 4.78 MeV.
The liquid scintillator was based on Eljen-309 which produces a larger scintillation tail for nuclear recoils versus electron-like recoils~\cite{PROSPECT:2019gwi}.
This property enables 'Pulse Shape Discrimination' (PSD) to classify event types.
PSD is calculated from digitized PMT waveforms, sampled at 4 ns intervals. Each pulse’s baseline is determined from pre-maximum samples, and the pulse area is integrated
from 3 samples before to 25 samples after the pulse maximum. The PSD value is defined as the ratio of the `tail' area (from 11 to 50 samples after the pulse arrival time)
to the total area (from 3 samples before to 50 samples after the arrival time), using trapezoidal interpolation between samples.
Window choices were optimized for neutron–gamma discrimination. When combined with energy information, the PSD provides a clear separation
of neutron capture events from positron interactions, crucial for identifying true IBD events among backgrounds such as fast neutrons from
cosmic rays and accidental coincidences.

During the course of the experiment, a subset of the PMT bases failed because of the ingress of liquid scintillator into the PMT housings, which caused damage to the voltage divider circuitry.
Segments with one or both PMTs failed were ignored for the initial IBD analysis~\cite{PROSPECT:2020sxr}.
Approximately a third of the segments were affected by the end of the run.
This naturally led to the question of whether it is possible to utilize information from single-ended (SE) segments, i.e those containing only one active PMT, for the purpose of position, energy and pulse shape (PSD) determination as opposed to double-ended (DE) segments.
Detailed studies of SE event reconstruction (SEER) were performed and a large improvement in background rejection achieved by extracting additional particle identification information~\cite{PROSPECT:2022wlf, PROSPECT:2024gps}.
This paper explores the possibility of reconstructing these quantities in SE segments using machine learning (ML) techniques.
The reconstructed quantities are then used for the purpose of background rejection.
A performance comparison is made between the ML-based analysis and a simpler approach based on the pulse shape of a single PMT.
Additionally, a comparison between the performance of different ML algorithms will also be shown.
We find that using ML improves the accuracy of the position, energy, and particle type reconstruction for single-ended events.
Finally, the impact of applying ML-based SE event reconstruction (SEER) to the IBD selection is evaluated.

\section{Single-Ended Event Reconstruction} \label{sec:SEER}
In this Section, we describe the methodology for reconstructing quantities of interest for an IBD analysis for SE segments.
First, a general description of the problem along with the conventional method for event reconstruction is given.
This is followed by a discussion on the utilization of machine learning techniques to improve upon the SE event reconstruction.
Finally, we demonstrate an improvement in our ability to reconstruct the position, energy, and particle type of the SE event.

\subsection{Description}
\begin{figure}[htbp]
	\centering
	\includegraphics[width=1.0\linewidth]{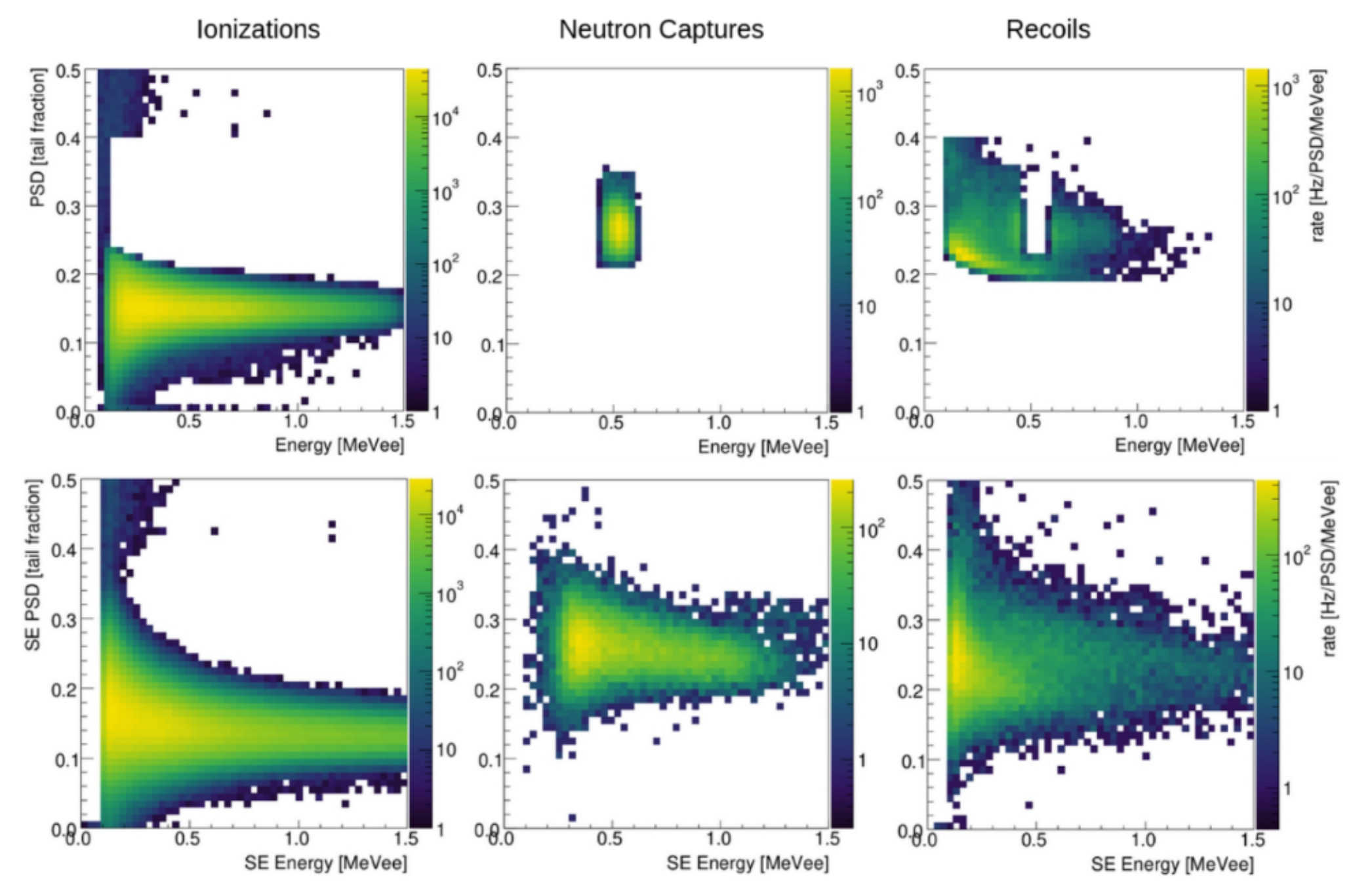}
	\caption{Events from a data taking period consisting of one day of
		reactor-on data split into electron-like ionization (left), neutron capture
		(center), and nuclear recoil (right) event classes.
		Top row:  DE reconstructed energy and PSD distributions.
		Bottom row:  SE reconstructed E and PSD distributions.
		The SE variables are calculated by ignoring the PMT that failed later in the experiment.
		The nominal DE selection is first applied to classify the events; the bottom row shows the SE variables for the same events.
	}
	\label{fig:E_PSD_SE_compare}
\end{figure}

As previously mentioned, the timing difference and relative signal amplitudes between PMTs in a segment are critical for reconstructing the position along the segment.
The timing-difference resolution between the PMTs corresponds to an RMS position resolution in $z$ of approximately 5~cm, relative to the 1.2~m segment length~\cite{PROSPECT:2020sxr}.
With only one PMT operational, the ability to reconstruct position diminishes greatly.
This significantly affects the energy estimation of the event due to the fact that a factor of three difference is seen in light collected between PMTs for events occurring near the PMT face.
On the other hand, the $z$ impact on the PSD is significantly smaller than its impact on energy, making the SE PSD a useful quantity for determining electron-like vs heavier recoiling particles.
This is illustrated in Figure~\ref{fig:E_PSD_SE_compare}, in which detector events are categorized into three classes as defined below.

(1) \textbf{Ionization} (electron-like) events are primarily due to high-energy gamma rays from reactor-related
backgrounds---such as neutron captures, high-energy beta decays, or muon-induced processes---that penetrate
the shielding and interact in the liquid scintillator, producing electrons via Compton scattering or pair production.
Additionally, reactor and cosmic neutrons interacting within the shielding and detector can create secondary processes
that yield gammas or electrons, also contributing to this category. Most notably, IBD positrons fall into this category.
(2) \textbf{Neutron capture} events correspond to neutron absorption on $^{6}$Li dopant in the scintillator,
producing a distinct signal from the $n + ^6\mathrm{Li} \rightarrow \alpha + t$ reaction. This constitutes the `delayed' signal of an IBD event, in addition to any other process that produces neutrons that thermalize within the detector.
(3) \textbf{Nuclear recoils} are primarily due to fast neutrons scattering elastically off protons or carbon nuclei in the scintillator,
producing a pulse shape with more light in the tail of the pulse (larger PSD) than electron-like ionizations.
%

%The selection is done by fitting gaussians to determine the mean ($\mu$) and standard deviation ($\sigma$) of the ionization events at various energy bins. An exponential is then fit to $\sigma$ and used to generate a selection window at $n \sigma$ above the mean. Events below this value are classified as ionizations and above recoils. The neutron capture window is also achieved via gaussian fits to the neutron capture event distribution in energy and PSD space, and a box around the mean is drawn inside of which events are considered neutron captures. For this illustration $2.0 \sigma$ widths were chosen. %
Energy is estimated for single-ended segments by simply assuming both PMTs received equal amounts of light, which is equivalent to assuming the event occurred at the center of the segment ($z$ = 0).
This creates a large increase in the spread in energy between the DE and SE variables.
Electron-like and nuclear-recoil-like bands are still separated, albeit less so.
Notably, while the total number of events is the same in both Double-Ended (DE) and Single-Ended (SE) panels of Figure~\ref{fig:E_PSD_SE_compare}
(aside from possible outliers falling outside the plotted window), the lower rate scale in the SE panel reflects the broader spread of events across
more bins due to reduced energy resolution. As a result, each bin contains fewer events, but the overall event count is conserved between panels.
The reader is referred to Reference~\cite{PROSPECT:2020sxr} for details on the calculation of the energy, PSD, timing, and position estimates, and to Reference~\cite{PROSPECT:2022wlf} for discussion of the conventional SEER implementation used for later PROSPECT physics analyses.

Light collection varies by roughly a factor of three over the length of the segment because each optical segment is 1.2\,m long, so scintillation light produced at one end undergoes
significant attenuation before it reaches the PMT at the opposite end.
This strong dependence on event position is why averaging the signals from both end PMTs is important for accurate energy reconstruction,
and it also explains the difficulty of extracting reliable energy information from a single-end readout.
Figure~\ref{fig:E_PSD_SE_compare} shows how this results in a large overlap between neutron capture and recoil signals when using SE energy and PSD, making neutron capture identification impossible.
The poor energy resolution makes event reconstruction for IBD interactions occurring within SE segments too unreliable to be useful for additional IBD statistics.
On the other hand, since the ability of the SE PSD to distinguish electron-like and nuclear recoils is only modestly degraded, this can be utilized for improved background rejection.

In order to distinguish electron-like recoils from nuclear recoils in SE segments, a Gaussian is fit to the electron-like PSD band for seven energy ranges between 90 keV
and 10.0 MeV.
The mean ($\mu$) and standard deviation ($\sigma$) of the fits are recorded.
The average of the means is stored in the calibration database for each segment.
A two parameter function is fit to the $\sigma$ distribution as a function of energy, $\sqrt{a^2 + b^2/E}$.
The best fit across all SE segments is used for the "SE PSD" cut, which checks if the SE PSD is above $\mu_i + n_{\sigma} \sqrt{a^2 + b^2/E}$, where $\mu_i$ is the segment $i$ mean SE PSD and $n_{\sigma}$ is the number of sigma above the mean for the cut threshold.
It was found that an $n_{\sigma}$ of 3.5 was best for maximizing the effective statistics of the dataset.

A veto window is also constructed around events containing any segment with SE PSD above the electron-like band.
A window of 170 $\mu$s after the detected recoil-like SE hit was found to optimize effective statistics of the dataset.
Any recoil-like delayed candidates in this window are rejected from the analysis.
This is called the "SE-recoil veto".

For IBD event selection, a prompt electron-like cluster of signals followed by a delayed neutron-capture-like signal is required.
Detector signals are grouped with a clustering algorithm using a 20 ns maximum time separation between pulses within the cluster.
We require that prompt candidate clusters must contain only ionization-like pulses in all SE and DE segments using their respective SE or DE PSD distributions.
Additionally, because we are only interested in positrons which deposit their energy in a DE segment, the SE segments within the prompt
cluster must have energies that are less than a threshold based on the possible SE energy of 511 keV gammas scattering from the positron annihilation.
This threshold was found to be 0.8~\mevee~of SE energy to optimize the effective statistics of the dataset.
We typically expect a maximum of around 0.5~\mevee, however, the SE energy has a large spread due to the lack of position information.
An improved estimate of the event position would lead to better energy estimation, and in turn, better IBD selection.
The machine learning (ML) techniques used for position reconstruction are described in the next section.

\subsection{Dataset} \label{sec:dataset}
The dataset used for training and validating the neural networks is depicted in Figure~\ref{fig:E_PSD_SE_compare}.
It is constructed by taking the first day of reactor-on data-taking and classifying the events according to the nominal IBD analysis~\cite{PROSPECT:2020sxr}.
The first day is chosen because all PMTs but one were operational at that time.
The classification of events follows established criteria: ionization events are identified by electron-like energy depositions
(0–15 MeV, PSD within 2$\sigma$ of the e-like mean); neutron captures by energy and PSD consistent with $n$-Li capture
(0.526 MeV, PSD within 2$\sigma$) and nuclear recoils are identified as events above the ionization PSD and outside the energy window of the neutron capture.

The dataset used for training and validating the neural networks is depicted in Figure~\ref{fig:E_PSD_SE_compare}.
It is constructed by taking the first day of reactor-on data-taking and classifying the events according to the nominal IBD analysis~\cite{PROSPECT:2020sxr}.
The first day is chosen because all PMTs but one were operational at that time.

The purity of event classification is a crucial consideration for training. As illustrated in Figure~\ref{fig:E_PSD_SE_compare}, the separation between the event classes in double-ended segments allows the selection of high-purity training samples. 
Most importantly for this study is the distinction between ionizations and other events. The ionization sample using the DE classification is well separated from the recoil and neutron capture events, giving a nearly 100\% purity sample for training purposes. Although cross contamination between recoil and neutron captures with the DE classification technique does occur, it is not as important for this work since the classifier is only used for rejecting non-ionization-like backgrounds.

Training for the neural networks is accomplished by ignoring the data for the PMTs that failed later in the run and using the remaining information as inputs to the network.
More details on training and validation can be found in Section~\ref{subsec:TrainingAndOptimization}.

\subsection{ML Based SEER Analysis} \label{sec:ML-SEER}
Machine learning is leveraged for SEER by predicting the energy and $z$ position of the SE event to improve IBD event selection.
One such improvement comes from the fact that positron events identified in DE segments tend to have low energy gammas in neighboring SE segments from the scattering of 511~keV annihilation gammas.
With a better energy estimate from ML, one can more accurately reject backgrounds where the neighboring energy depositions are too large to actually be annihilation gammas.

We also use ML to create a classifier that gives a class score for each segment that detected energy during an event.
The score represents the likelihood that signal came from an event of the given class.
This information can be used for particle identification (PID).
In the \pspt~detector, five major event classes were assigned for the purpose of IBD selection.
These are labelled electron-like (ionization), proton-like (recoil), neutron capture on \el{Li}{6} (ncap), muon, and ingress.
Any signal detected, whether from background or IBD events, is classified into one of these categories.
The events are classified using DE information in the manner described in the previous section for ncap, ionization, and recoil.
Muons are any event with a total energy above 15 MeV.
Events above the recoil-like PSD are considered ‘ingress’ events, which occurred within scintillator that leaked into the PMT housings.
These have very high PSD, destroying the ability to reconstruct the particle type, so the event is discarded.
A summary of the different event types is found in Table~\ref{table:PID}.

The next two sections will give details on how convolutional neural networks and graph convolutional networks were utilized to improve SEER.
In each section we will describe the algorithms used, preparation of the waveforms, dataset creation, training and optimization of the network weights, and give details on the performance of the networks for predicting SE position, energy, and classification.

\begin{table}[htbp]
	\centering
	\caption{Table of different particle identification types and their
		descriptions.}
	\smallskip
	\label{table:PID}
	\begin{tabular}{|l|l|}
		\hline
		\bf Class       & \bf Description
		\\
		\hline
		Ionization      & Events inside the electron-like ionization
		PSD band                                                        \\
		Recoil          & Events inside the nuclear recoil band
		\\
		Neutron Capture & Events within the neutron capture PSD and
		energy range                                                    \\
		Muon            & High energy events (total visible energy > 15
		MeV)                                                            \\
		Ingress         & Events with PSD above the neutron capture
		band                                                            \\
		\hline
	\end{tabular}
\end{table}

\section{Convolutional Neural Networks (CNNs) for SEER} \label{subsec:CNN}
Convolutional neural networks (CNNs) as first introduced by \cite{FUKUSHIMA} are deep neural networks that are ideal for learning translationally invariant features of spatial data.
CNNs consist of several convolutional layers chained together.
Each convolutional layer applies a set of filters (also called kernels) to process the input data.
Each filter is a small matrix of trainable weights.
During the convolution operation, these filters systematically slide across the input data, segment by segment.
At each position, the filter performs an element-wise multiplication between its weights
and the corresponding receptive field of the input data.
The sum of these multiplications becomes a single value in the output feature map.
The output of the series of convolutional layers can then be fed into a linear network in order to output a classification score or some other metric.
For a mathematical review of convolutional network layers, we refer readers to Reference~\cite{CNN_OG}.

The convolution operation combines information across all input channels to produce each output channel (feature map).
Each output channel is produced by convolving all input channels with a separate set of trainable kernels and summing the results.
In the case of PROSPECT data, the input channels are the two PMT waveforms concatenated together, or the extracted features in the feature-based models.

In this work, the output of each convolutional layer is passed through a batch normalization layer, followed by an activation layer.
A batch is a subset of the training data that is evaluated together in one forward and backward pass, during which the gradients of all model parameters with respect to the loss are computed. These gradients are then used to update the model parameters according to the chosen optimizer.
The batch normalization layer~\cite{batchnorm} transforms the value at each channel in the output of the convolutional layer, $\hat{x}_k$, into the signed mean variance for that channel, $\bar{x}_k$:

\begin{equation} \label{eq:batchnorm} \bar{x}_k = \frac{\hat{x}_k - \mu_k}{\sqrt{\sigma^{2}_k} + \epsilon} * \gamma_k + \beta_k.
\end{equation}

The calculations of the mean, $\mu_k$, and variance $\sigma^{2}_k$, are performed over a single batch of events rather than the entire dataset.
The variable $\epsilon$ is a small number used to avoid division by zero, and $\gamma_k$ and $\beta_k$ are free parameters designed to scale and shift the output of the normalization.
Batch normalization layers are widely used in CNNs for the purpose of speeding up training of the network parameters.
The activation layer used is a rectified linear unit (ReLU) layer which sets negative values of the input to 0.

To apply CNNs on PROSPECT data, one must choose how to map the PMT data to the inputs of the CNN.
The natural choice given the spatial arrangement of the detector segments is to assign the features of each of the 14x11 segments to a spatial input.
In this scenario, the features fed into the model are either the waveforms of each of the two PMTs concatenated as described in Section \ref{subsec:waveform-preparation} or the extracted features as described in Section~\ref{subsec:extracted-quant}.
Signals are clustered in time with a 20 ns maximum separation time between pulses~\cite{PROSPECT:2020sxr}, and each segment containing PMTs with signals above threshold corresponds to the inputs to the CNN.
All segments that did not detect pulses above threshold in a given cluster are ignored.
Note that this cluster of signals defines an `event' for CNN input purposes, with all pulses within this cluster processed together as a single input instance.
Because the typical event occurring in the PROSPECT detector only contains 2 -- 4 segments with signals above threshold, the input to the neural network is very sparse.
This makes the PROSPECT data ideal for sparse convolutional networks using sparse matrix algebra to calculate the output of each convolutional layer.
For this reason, the sparse CNN library spconv~\cite{SPCONVp} was chosen.
The spconv library allows one to use submanifold sparse convolutional layers~\cite{SUBMANIFOLD} which only transfers information from nonzero segments to the next layer.
This technique significantly reduces the computational burden of the network without any loss of information relevant for segment level predictions.
All CNN networks used for SE reconstruction utilize these submanifold convolutions for every convolutional layer.

\subsection{Data Preparation for CNNs}
There are two types of inputs for the CNN models:  (1) PMT waveforms after baseline subtraction, time alignment and value normalization is applied, and (2) extracted features from the waveforms and DE reconstruction when applicable.
Henceforth when referring to ``waveform'' models or ``extracted feature'' models, we are referring to models that either utilize the full waveform from the PMT or the extracted features.
The preparation of these inputs are described in detail below.
Analog waveforms from each PMT are digitized by a CAEN V1725 250 MHz 14-bit waveform digitizer.
We will refer to each value of the digitized signal as a sample which represents the average voltage over a 4 ns sampling interval.
While the clustering algorithm groups pulses within clusters separated by no more than 20 ns between subsequent pulses' arrival times,
the CNN input for each waveform is standardized to 65 samples (260 ns) per segment.
This extended sample window enables capture of the full pulse shape and timing offsets between segments within a cluster, allowing
for clusters with signals of varying start times to be fully captured and properly time-aligned as described in the next section.

\subsubsection{Waveform Preparation}\label{subsec:waveform-preparation}
First, waveforms are clustered based on a maximum separation time of 20 ns and the pulses are aligned in time using the calibrated arrival times relative to the event trigger.
65 samples are allocated for each waveform.
During early PROSPECT R\&D a study optimizing the PSD metric found that 54 samples were needed to best discriminate ionization-like pulses from nuclear-recoil-like pulses.
An additional 5 samples were used for the purposes of this study and 6 samples were added for potential offsets between detectors in a given cluster.
The first pulse that arrived is set so that the eighth sample contains the arrival time.
All other pulses are offset such that their arrival sample is at sample n, where $n = \left(\textrm{round}((t_{i} - t_{0})/4) + 8\right)$, $t_{i}$ is the arrival time of the $i\textrm{th}$ waveform in the cluster and $t_{0}$ is the arrival time of the first waveform.
The round function selects the nearest integer, rounding down for midpoint values.

Once the waveforms are time-aligned, the samples are scaled by the inverse of the PMT gain to regularize their magnitudes.
Before feeding these values into a neural network they are scaled to 32-bit floating point numbers between 0 and 1 by dividing by $2^{14} - 1$.

\subsubsection{Extracted Features Inputs}\label{subsec:extracted-quant}
A natural choice for inputs into the neural network are the quantities extracted from the waveforms used for the physics analysis.
These are the pulse area, PSD and arrival time of each of the waveforms in a cluster.
Two key aspects of CNNs warrant investigation: first, how their performance using only these extracted variables compares to using complete waveform data; and second, whether extracting additional features from the waveforms (beyond those currently used in IBD analysis) could help bridge any performance gap between the two approaches.

In this work we chose to utilize the pulse area, arrival time, and PSD for each live PMT.
Data is grouped by segment, so each pair of PMT features are concatenated together.
If both PMTs are alive in a segment, the reconstructed energy, position, segment axis position, and pulse shape are also included.
Data is normalized using scaling factors chosen to map typical signals between 0 and 1.
While some values exceed this range, this preserves important outlier information while maintaining stable training through batch normalization and activation functions.
Features not available for a given segment are set to 0.

A second model based on extracted features was also utilized that contains information not used for the IBD analysis.
These additional features are the pulse width, rise time, fall time, and pulse height.
It was found that the addition of these variables improved the model compared to using only features relevant to the IBD reconstruction analysis, although not to the point of being at parity with the full waveform model.
A full model comparison is shown for SE $z$ reconstruction performance in Table~\ref{table:ZPerformanceComparison}.

\subsection{Core CNN Architecture for SE Reconstruction}
\label{subsec:CoreCNNArchitecture}
The core convolutional neural network (CNN) architecture utilized throughout this work serves as a shared engine for SE segment-level tasks,
including $z$ (position) reconstruction, energy reconstruction, and PID classification. The input to the network is a 14$\times$11 array of
segments, with each segment containing 130 input features given by concatenated waveforms from the two PMTs. Generic convolutional layers with
filter sizes of 1, 3, 5, and 7 are used, and the number of convolutional layers is a tunable hyperparameter
(parameters describing the architecture of the network), varied up to 12 layers.
Padding is set such that the output shape is preserved as 14$\times$11. This allows for segment-level predictions such that the final output represents the
predicted quantity for each segment. Expansion and contraction of feature sizes, driven by additional tunable
hyperparameters, enable the model to increase and then reduce the number of features at each layer to the required output dimension for each task—either a
single value (e.g., $z$ or energy) or, for classification, a vector of class scores. All but the final layer are followed by
batch normalization and ReLU activation.

The hyperparameters were varied using a gradient-free optimization method as described in Section~\ref{subsec:TrainingAndOptimization}.
Through experimentation it was found that layers with larger numbers of intermediate layers outperformed smaller numbers up to around 10 layers, beyond which there was no improvement.

\subsection{Position Reconstruction Using CNNs} \label{subsec:CNNPos}
Position reconstruction is a difficult problem for SE segments because the nominal position reconstruction algorithm relies on both the light ratio
between PMTs and the differential timing between PMTs.
Any SE segment reconstruction method must rely solely on the timing distribution of the light collected for each pulse within the cluster.
Both full waveforms and extracted quantities are used as inputs to the CNN core architecture (see Section~\ref{subsec:CoreCNNArchitecture}).
The results were benchmarked against a simple "nearest neighbors average" approach in which we simply average the z position reconstructed from neighboring double-ended segments.
The architecture of the best network found on a dataset containing all types of events that could be seen in an IBD analysis and the full relevant energy range (0.05 - 10.0 MeV SE energy) can be found in Table~\ref{table:BestZModelDiagram}.

\begin{table}[htbp]
	\centering
	\caption{This table lists the shape of each convolutional kernel used
		for the best performing $z$ reconstruction.
		Filter size is the dimensions of the convolutional kernel.
		A 1×1 filter processes each point individually, while larger filters (3×3, 5×5, 7×7) analyze patterns across neighboring detector segments.
		Input size represents the number of features at each node of the network at that layer.
		The 130 features at the first layer are the concatenated waveforms from the two PMTs for each segment.
		Output size represents the number of resulting features after the convolution is performed, which is equal to the number of convolutional kernels used at that layer.
	}
	\smallskip
	\label{table:BestZModelDiagram}
	\begin{tabular}{|l | l | l | l|}
		\hline
		\bf Layer Number & \bf Filter Size & \bf Input Size & \bf
		Output Size                                               \\
		\hline
		1                & 1$\times$1      & 130            & 310
		\\
		2                & 1$\times$1      & 310            & 286
		\\
		3                & 1$\times$1      & 286            & 262
		\\
		4                & 1$\times$1      & 262            & 238
		\\
		5                & 1$\times$1      & 238            & 214
		\\
		6                & 1$\times$1      & 214            & 190
		\\
		7                & 1$\times$1      & 190            & 166
		\\
		8                & 7$\times$7      & 166            & 142
		\\
		9                & 5$\times$5      & 142            & 118
		\\
		10               & 3$\times$3      & 118            & 94
		\\
		11               & 3$\times$3      & 94             & 70
		\\
		12               & 3$\times$3      & 70             & 46
		\\
		13               & 1$\times$1      & 46             & 22
		\\
		14               & 1$\times$1      & 22             & 1
		\\
		\hline
	\end{tabular}
\end{table}

\subsection{Energy Reconstruction Using CNNs}
\label{subsec:EnergyReconstruction}
SE energy reconstruction with CNNs is treated similarly to segment-level $z$ reconstruction, employing the same core CNN architecture described in Section~\ref{subsec:CoreCNNArchitecture}.
The $z$ distribution of events is roughly linear, with backgrounds slightly favoring one side due to ambient gamma backgrounds in the experiment hall, whereas the energy distribution exponentially decays at higher energies due to the increasing rarity of processes producing higher energy gammas.

\begin{figure}[htbp]
	\centering
	\includegraphics[width=0.6\linewidth]{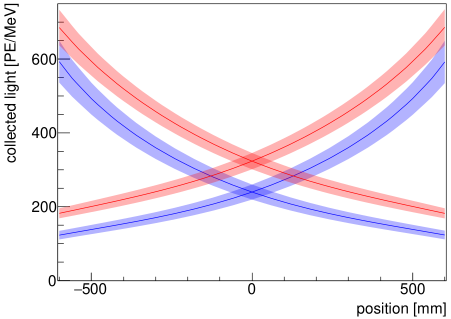}
	\caption{
		Light collection curves measured from data, averaged over all channels, for neutron capture events.
		The upper red curves represent the light collection at the beginning of data taking (averaged over a two-week period), while the lower blue curves show the averages at the end of the data set (final two-week period).
		Bands indicate the RMS spread between channels.
		These curves illustrate the measured time-dependent variation in the position-dependent light collection used for calibration. Data and method details are described in~\cite{PROSPECT:2020sxr}.
	}
	\label{fig:lightcollection}
\end{figure}

An alternative method to using the network to directly predict the true energy is to use the $z$ prediction from the neural network and calculate what the energy would be given that position.
Calibration curves representing the relative average light collected between the two PMTs for events occurring at different positions along
the segment are used to predict the light that would have been seen by the other PMT if it were working, then the combined light output is used to calculate the position-corrected energy prediction.
A plot of the light collection curves taken by averaging the light output as a function of position from neutron capture events for each segment over two week time periods can be found in Figure~\ref{fig:lightcollection}.

Through the best hyperparameter optimization on our training dataset (see ~\ref{subsec:TrainingAndOptimization}) it was found that the error in energy predicted as a function of true energy tended to match the best predictions based on inverse-calibration using the best $z$ prediction models at energies greater than 0.5 MeV, but for lower energies the $z$ prediction based model outperformed.
For this reason, we chose to utilize the $z$ prediction model to predict the energy rather than having a dedicated neural network model for energy prediction on single-ended segments.

\subsection{PID Classification with CNNs} \label{PIDCNN}
PID is reconstructed using the same core CNN architecture, with the only difference being that the output contains five quantities per segment which we call class scores,
representing the likelihood of each respective PID class described in Section~\ref{sec:ML-SEER}. The network thus terminates in a softmax layer,
with the cross-entropy loss minimized as in Equation~\ref{eq:crossentropyloss}.
The truth value that the model is trained on is the category given by the dual ended analysis.
Cross-entropy loss is commonly used for classification problems.
For a problem containing $C$ classes, It is defined for each sample as

\begin{equation} \label{eq:crossentropyloss} {\rm loss}(x, j) = -\log\left(\frac{\exp(x_j)}{\sum_{i=0}^{C-1}\exp{(x_i)}}\right) \end{equation}

where $j$ is the index of the correct class and $x$ is the vector of class scores for the sample.
%\begin{equation} \label{eq:crossentropyloss} {\rm loss}(x) = -\log\left(\frac{\exp(x_j)}{\sum_{i=0}^{C-1}\exp{x_i}}\right). \end{equation} 

\subsection{CNN Training and Optimization}
\label{subsec:TrainingAndOptimization}
Training is done using stochastic gradient descent with Nesterov momentum as first described by \cite{nesterov}.
Typical values used for the momentum are 0.90 -- 0.98 and learning rates around 0.01 -- 0.02.
An ExponentialLR learning rate scheduler is applied during training, with a decay factor $\gamma = 0.9$, so that the learning rate decreases multiplicatively after each epoch.
PyTorch's automatic differentiation algorithm is used to calculate gradients in batches that are around 10,000 to 20,000 events in size~\cite{PYTORCH}.
The PyTorch Lightning framework as detailed in \cite{LIGHTNING} is used to organize the ML code in a way that leverages best-practices recommended by professionals in the ML community.
These include separation of research code from engineering code, standardized model training loops, and built-in logging and experiment tracking.

Separate subsets of the available data are used for training, validation, and testing.
Typically the validation dataset is roughly 1/3 to 1/5 the size of the training dataset.
Depending on the phase space sampled for the training set, the number of samples used for training is typically between 1 -- 3 million events.
It was found experimentally that training sets using less than 1 million events tended to perform worse, but using more than this gave negligible performance benefits.

The truth data used in training is the calibrated physics quantity found for double-ended readouts.
Because the PMTs gradually failed throughout the several month long data taking period, training and validation data sets were constructed from the first day of data in which all PMTs were operational, and calibrated physics quantities from PMTs could be utilized to train the single-ended prediction models.
The dead PMTs were simulated by ignoring their waveforms.

Training is split into epochs.
An epoch consists of one full pass of the training dataset in which the gradients of each batch are computed and model weights are updated until all batches in the dataset have been exhausted.
At this point the model is tested on the validation dataset.
This sequence constitutes a single epoch.
The loss for the validation dataset in a given epoch is compared to the current best loss of all epochs.
If the loss fails to improve by a certain amount for a number of epochs in a row, the training is automatically ended.
This criterion varies from problem to problem, but typical values used in this work were five epochs failing to improve the loss by 0.01 resulted in cancellation of the training run.
Model weights for the best validation loss are saved for later use.
Splitting into separate training and validation datasets in this manner prevents over-training in which the model finds patterns in the training dataset that don't generalize to other data.

Typically, given the large size of the training sets, the training usually only needs to run for 30 or so epochs before reaching the limit of its loss reduction.
The loss function used is the mean absolute error for scalar quantities like $z$ position or energy, and cross-entropy loss for classification tasks.
We have also tested mean squared error (MSE) as a loss for these regression tasks, but observed no significant difference in the resulting performance.
Cross-entropy loss is defined in equation~\ref{eq:crossentropyloss}.

The models used will generate outputs for each segment, but we only care about the performance of the model for SE segments.
It was found that training only based on the loss for SE segments found optimal performance for these segments and faster convergence.
Thus, for training we ignore the loss on DE segments.

Optimization is performed using a gradient-free optimization library Optuna~\cite{OPTUNA}.
Hyperparameters of the model are chosen and varied within bounds that are allowed given the memory constraints and CPU resources available.
A typical set up for the hyperparameter search would be to set the number of trials to 100, vary between three and five hyperparameters, and set a flag to prune unpromising trials.
It was found that the default hyperparameter sampler and trial pruner was sufficient for our needs.
The default sampling algorithm is the Tree-structured Parzen Estimator algorithm~\cite{TPE}.
The default pruner is the median pruner, which will prune the trial if the trial's best intermediate result is worse than median for previous trials at the same epoch.

A sqlite~\cite{sqlite2020hipp} database was used to store trial results.
This allowed for easy parallelization of the hyperparameter optimization process.
Multiple hyperparameter runs were set off on a process using two threads, one thread reserved for data computations and one for data loading.
Each run would be allocated approximately two days of processing time.
Each thread references the same database for determining loss values for previous trials used for the next parameter selection and as a reference for when to prune unpromising trials.
We found that when using a CPU cluster for hyperparameter optimization, only using one thread for computations and one for data loading was by far the most efficient choice, as increasing this number only gave marginal gains in terms of computational efficiency.
For example, using two threads for computations sped up the time for a single epoch in one test by 30\% at the cost of double the resources.
With only two threads in use for training, we could run many more models simultaneously with the available CPU resources.

\subsection{CNN Performance Summary}
\begin{figure}[htbp]
	\centering
	\includegraphics[width=1.0\linewidth]{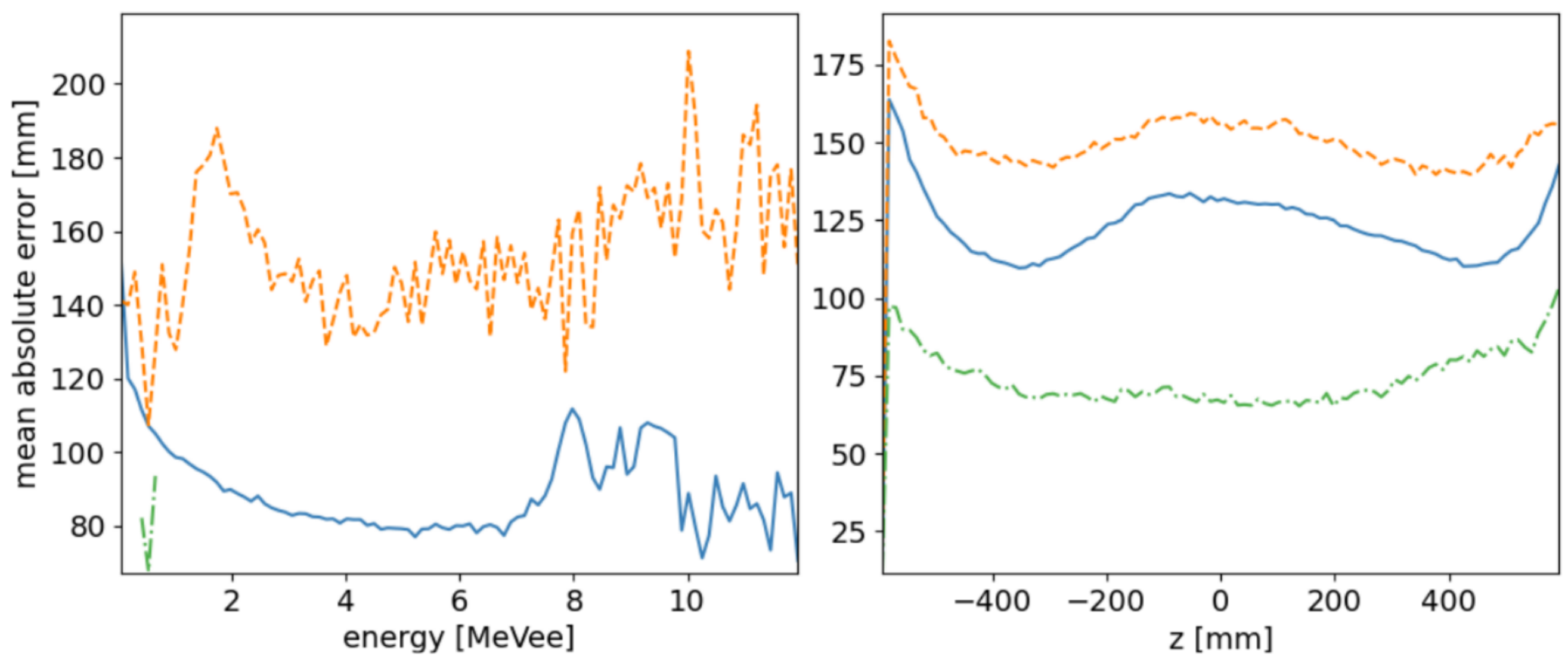}
	\caption{Mean absolute error (MAE) of the reconstructed $z$ as a function of energy (left) and $z$ (right) for different event classes.}
	\label{fig:ZPerformance}
\end{figure}

\begin{figure}[htbp]
	\centering
	\includegraphics[width=0.6\linewidth]{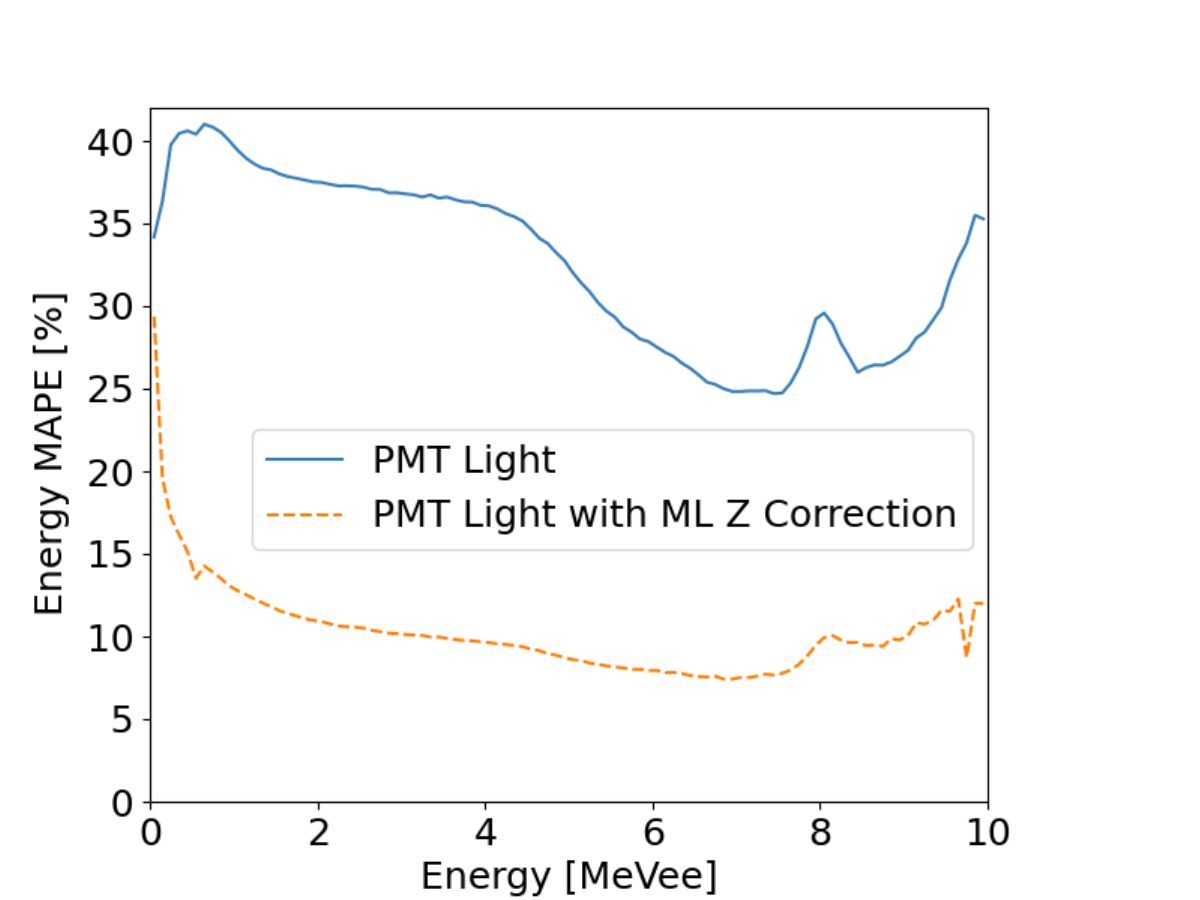}
	\caption{Mean absolute percentage error (MAPE) on reconstructed energy
		reconstruction performance using the $z$ prediction.
		Low energies ($<$ 0.1 MeV) perform poorly due to the algorithm's poor $z$ reconstruction ability at low light levels.
		Larger errors at high energies ($>$ 7.8 MeV) are due to less training data available at this region of the phase space.
		Blue curve is calculated assuming equal light in both PMTs ($z$ = 0).
	}
	\label{fig:MAPE}
\end{figure}

The best performing CNN model for position reconstruction was shown in Table~\ref{table:BestZModelDiagram}.
This model achieved a mean absolute error of 133 mm on the first day of data used for the IBD analysis.
The mean absolute error as a function of energy and position is plotted in Figure~\ref{fig:ZPerformance}.
In this plot we see that the neutron capture class shows the best $z$ reconstruction performance due to the narrow range of light output typical for it.
The energy reconstruction performance using the $z$ prediction from this model is shown in Figure~\ref{fig:MAPE}, with a mean absolute percentage error of less than 10\% for most of the relevant energy range.
Note that the IBD analysis would cut any events occurring with the neutron capture energy and PSD bounds with a multiplicity greater than 1 since these are recoils that coincide with the neutron capture window, but for the purposes of illustrating performance we label all of the events within this window "neutron capture".

For PID classification, the best performing CNN was found to have a 10\% higher cross-entropy loss on the test dataset than the best graph network which will be described in the subsequent section.
For this reason no further detailed classification performance was investigated with CNNs.
Detailed classification performance can be found in Section~\ref{sec:GraphPerformance}.

\subsubsection{Position Scan Calibration Data}
In addition to testing on the first day of data, we tested the algorithm on position scan calibration data.
The data shown were taken with a $^{137}\textrm{Cs}$ source deployed at various locations along the length of the segment in one of the corner rod source deployment tubes~\cite{PROSPECT:2020sxr}.
Two runs were selected for comparison: one in April (shortly after the start of data-taking) and another in August.
Note that the April data had functioning PMTs that were later turned off due to issues previously mentioned, so this analysis ignores those data in the same way the neural network training is done ignoring the final set of PMTs that were non-functional.

\begin{figure}[htbp]
	\centering
	\includegraphics[width=1.0\linewidth]{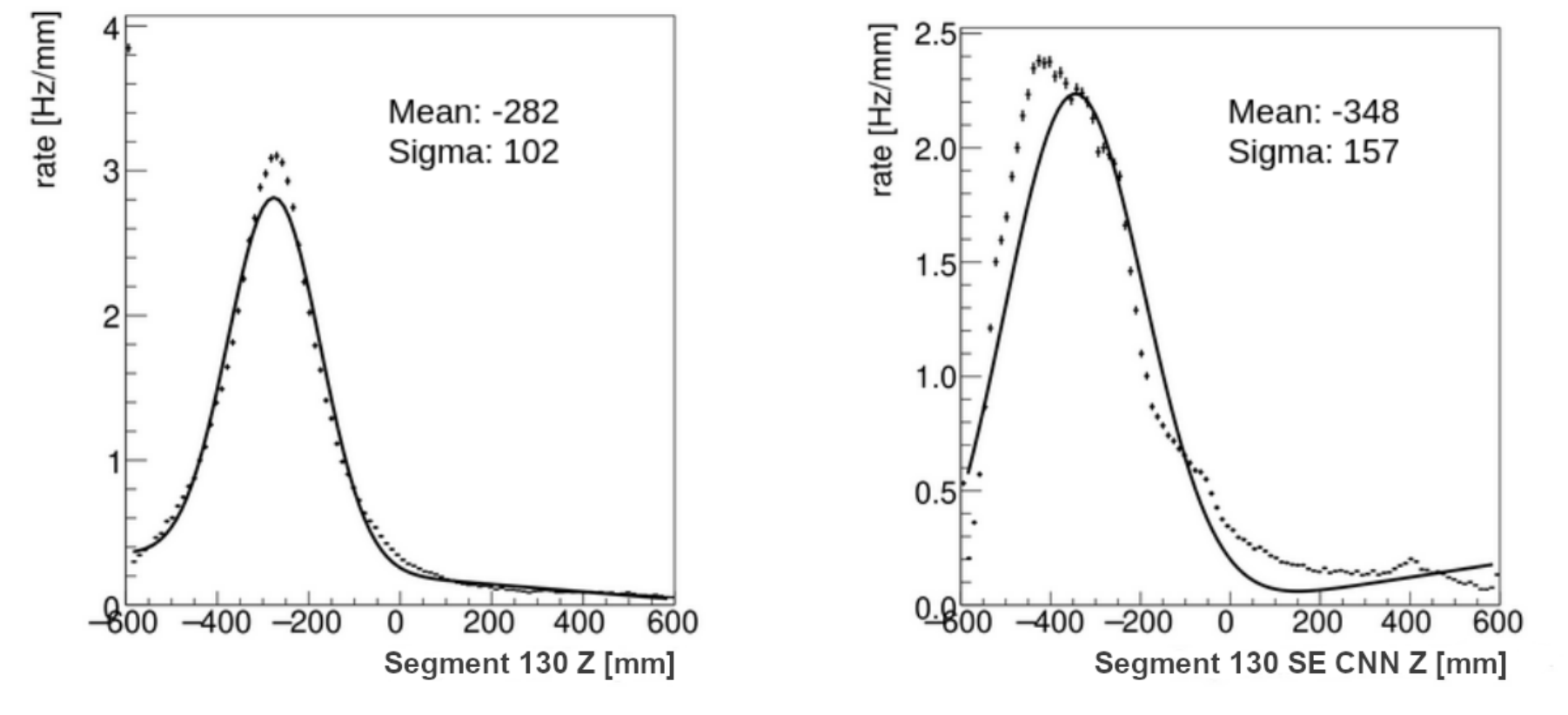}
	\caption{The DE $z$ reconstruction method from calibration is shown on
		the left.
		The CNN reconstruction method is shown on the right using only the waveform from a single PMT along with neighboring PMT information.
		In this run the source was placed at -300 mm.
		The solid line is the fit to the data.
	}
	\label{fig:ZCalibrationSample}
\end{figure}

\begin{figure}[htbp]
	\centering
	\includegraphics[width=1.0\linewidth]{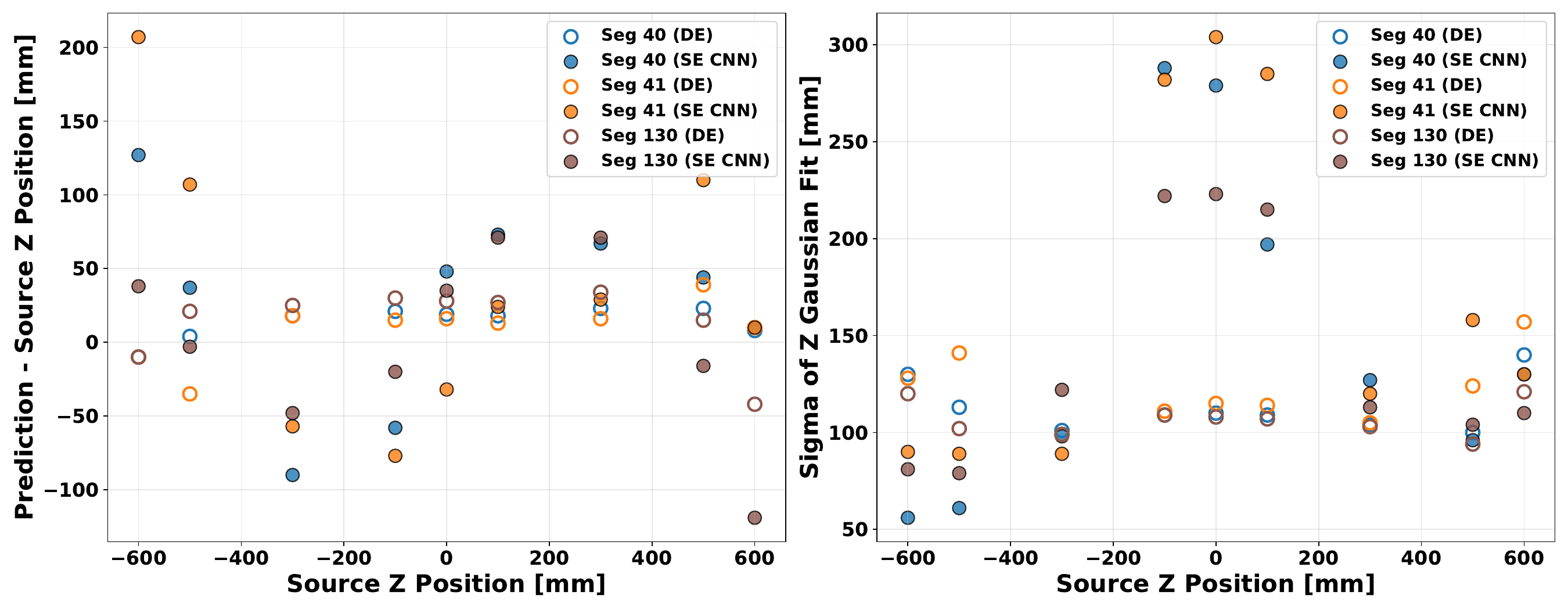}
	\caption{(Left) Fitted Gaussian mean minus true source position at 3 different
		segments for a calibration run performed in April 2018.
		The nominal DE position reconstruction are shown with open marks and SE CNN reconstruction are shown in filled marks. (Right) The associated fitted Gaussian sigma width for both methods.
	}
	\label{fig:pos_cal}
\end{figure}

Pictured in Figure~\ref{fig:ZCalibrationSample} is an example position reconstructed with the DE information and the ML SE reconstruction for comparison.
A Gaussian function summed with a linear background was fit to the data, and the mean of the Gaussian was compared to the actual known source position.
This was done for each source location within each day of runs.
The results for a full set of calibration runs performed in April are shown for three segments in Figure~\ref{fig:pos_cal}.
This shows that ML reconstruction can predict location, albeit with reduced accuracy.
It is important to note that the traditional analysis relied on relative amplitude and timing between PMTs to make a $z$ prediction, so this reduced accuracy is expected.
Both methods lose accuracy over time due to degradation of scintillator light yield, as described in~\cite{PROSPECT:2020sxr}.

%\begin{table}[htbp]
%\centering
%        \caption{Results from a sample of the calibration data. \textmd{The difference between calibration source position and reconstructed position is calculated with a gaussian fit to the reconstructed position histogram and the average absolute difference between source positions and centroid is displayed for each segment. It is important to note that the ML prediction is made using only data from one PMT along with neighboring information from working PMTs, so it is expected to be worse than the standard $z$ prediction algorithm which relies on both PMTs being functional.}}
%        \smallskip
%        \label{table:ZCalibrationResults}
%        \begin{tabular}{|l |l| l| l|}
%        \hline
%           \bf Calibration Run & \bf Segment & \bf DE Cal Difference [mm] & \bf SE ML Difference [mm] \\
%        \hline
%            April & 40 & 16 & 61 \\
%            April & 41 & 19 & 73 \\
%            April & 130 & 26 & 47 \\
%            September & 130 & 21 & 68  \\
%        \hline
%        \end{tabular}
%    \end{table}

\section{Graph Neural Networks (GNNs) for SEER} \label{sec:GCN}
\begin{figure} \centering
	\includegraphics[width=1.0\linewidth]{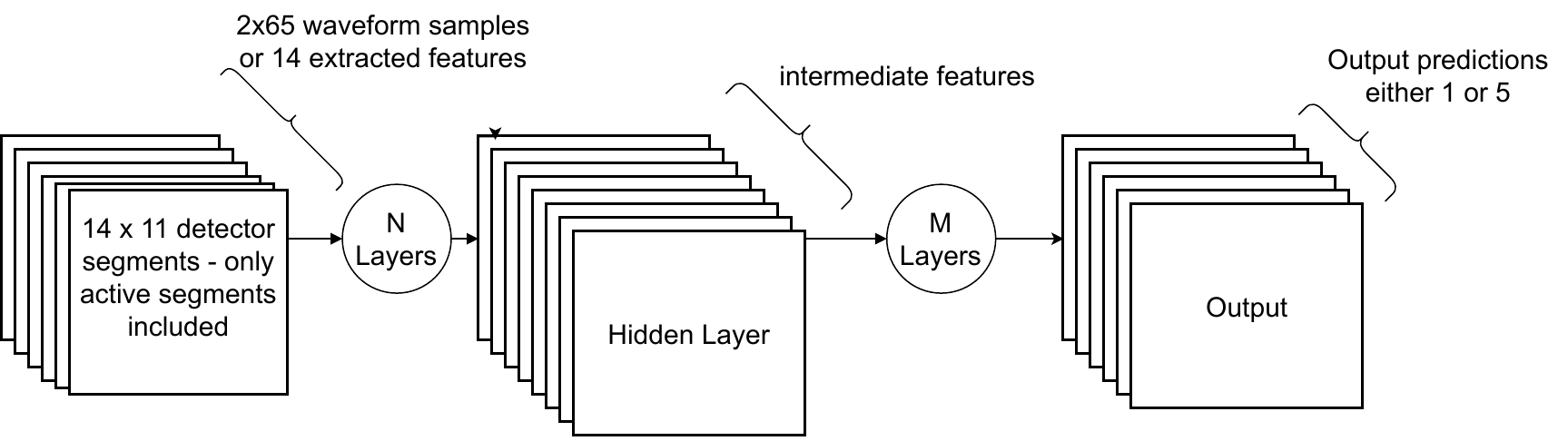}
	\caption{General architecture used for both $z$ and PID predictions for each
		energy deposition event.
		Note that only active segments are used as inputs and computations are all done using sparse matrix algebra (either sparse CNNs or graph convolutions).
		The typical pattern is for $N$ expansion and $M$ contraction layers, each of which linearly increases then decreases the number of features down to either five outputs for PID prediction or one output for $z$ prediction.
		In the case of CNNs, the intermediate layers can be 1$\times$1, 3$\times$3, 5$\times$5, or 7$\times$7 convolutions and in the case of graphs the convolutions depend on the type of GNN chosen and the choice of $k$ using the $k$-nearest-neighbors edge construction.
	} \label{fig:generalarch}
\end{figure}

GNNs have already been shown to be useful in particle physics experiments~\cite{GCNParticle} and various neutrino physics experiments, including IceCube~\cite{GCNIce}, liquid argon-based time projection chambers such as DUNE~\cite{GCNDUNE}~\cite{GCNTPC}, and scintillator-based neutrino experiments~\cite{GCNScint}.
GNNs are ideal for PROSPECT data due to their effectiveness at representing sparse datasets.
Graphs can be arbitrarily large and therefore code written for training on graphs implements sparse matrix algebra~\cite{PYTORCHGEOM}.
Implementation of the GNN on PROSPECT data follows the architecture modeled by the CNNs as described in Section~\ref{subsec:CNN}.
This general architecture is shown in Figure~\ref{fig:generalarch}.
Graphs are composed of nodes and edges.
Similar to the CNN construction, we utilize event clusters as individual graphs.
The `nodes' of the graph are segments that had one or more PMTs above threshold.
The features of each node are then either the two PMT waveforms concatenated together or the extracted waveform features.
The `edges' of the graph are defined using a $k$-nearest neighbors approach.
Here $k$ is a parameter of the network that represents the maximum number of connections a segment can have.
If an event contains more segments above threshold then the segment will be connected to the segments in closest proximity.
Through experimentation, it was found that a $k$ value of 4 or greater was optimal.
Since most of the events contain a multiplicity of 5 or less, this means that for most events each segment is connected to every other segment.

When processing a graph, GNNs update each node's feature vector by aggregating information from its neighboring nodes through message-passing operations along the graph edges.
This allows the network to learn meaningful representations based on both the local structure and node attributes.
Different GNN layer types implement various approaches to this aggregation step, potentially using attention mechanisms, edge weights, or learned functions.

In some graph network constructions, there are weights associated with each edge of the graph.
These weights modify the magnitude of the convolution operation.
In our testing, we found that graph networks that utilize edge weights outperform those without edge weights.
For PROSPECT, we used edge weights proportional to the Cartesian distance between segments.

\subsection{Types of Graph Networks Tested}
There are a number of ways GNNs can be implemented.
The python package PyTorch Geometric~\cite{PYTORCHGEOM} is utilized for the implementation and as of this writing has 42 different graph neural network (GNN) layer operators implemented.
Rather than completing a detailed study of each one of these operators to choose one that best fits our use case, a brute force approach was used where the easiest ones to implement were tested.
For this test we used a fixed architecture other than the specific GNN layer which was randomly varied over about 100 trials.
The problem these were tested on was the SE $z$ prediction.
The top performing operators were then subjected to further hyperparameter testing to determine which one worked best for $z$ prediction.

Of the 42 operators available in PyTorch Geometric, 18 different operators were implemented.
Of these 18, the 6 top performers were GCNConv~\cite{GCNKIPF}, FeaStConv~\cite{FEASTCONV}, SuperGATConv~\cite{SUPERGATCONV}, PointConv~\cite{POINTCONV}, GMMConv~\cite{GMMCONV}, and ARMAConv~\cite{ARMACONV}.
Of these, GMMConv was chosen due to it outperforming the others by a slim margin on a hyperparameter search. This search involved
varying the number of layers and the number of channels used at each layer. The number of channels was parameterized in the same manner as was described
in section~\ref{subsec:CoreCNNArchitecture}, except that the outputs shrink to 5 representing the different possible event classes.
Since GMMConv outperformed the others, we will focus on it for the remainder of this paper.

\subsection{Core GNN Architecture for SE Reconstruction}
\label{subsec:CoreGNNArchitecture}
The core graph neural network (GNN) engine described here underlies all GNN-based segment-level tasks, including PID classification
and $z$ position reconstruction. Each event is represented as a variable-sized graph with nodes for segments with above-threshold signals and
edge connections defined either by $k$-nearest-neighbors or by spatial adjacency. Node-level features (typically the concatenated PMT waveforms)
serve as inputs. The sequence of GMMConv layers is followed by batch normalization and ReLU activation, except for the final output layer,
which omits normalization and activation. Feature expansion and contraction is hyperparameter controlled. The output feature size is adjusted to
match the target for each task (i.e., one value for $z$ or five for PID).

The primary difference between the graph architecture and the equivalent CNN architecture is the choice of adjacency matrix which determines which segment features are convolved together.
The $k$-nearest-neighbors ($k$-NN) approach was utilized for PID classification and $z$ prediction.
This means that for each segment in the cluster, the $k$-nearest segments above threshold in the cluster are connected to it.
Self-loops can also be included, meaning the segment's own features can be convolved with the layer weights for that segment's output.
The choice to use self loops along with the parameter $k$ were varied in the hyperparameter search, and it was found that self loops are not necessary along with $k$ values of greater than or equal to 6.
Because most events have a multiplicity of 4 or fewer, it was optimal to connect all of the segment features with each other segment in the event.

Similar to the CNN architecture, we optimize the architecture using a parameter that represents the expansion of the feature size, the number of expansion layers, and the number of contraction layers.
We also varied the number of nearest neighbors for constructing the adjacency matrix between 2 and 8.
These parameters are all varied for a hyperparameter search as described in Section~\ref{subsec:TrainingAndOptimization}.
The best model found for PID classification is shown in Table~\ref{table:BestPIDGraphModel}.
This model had a $k$ value of 5 nearest neighbors and used the Cartesian distance between segments as the edge weights for each GMMConv layer.

GMMConv utilizes a Gaussian mixture model in its weights used in the convolutional layers.
The exact form of a specific layer is~\cite{GMMCONV}

\begin{equation} \label{eq:GMMConv} \mathbf{x}^{\prime}_i = \frac{1}{|\mathcal{N}_i|} \sum_{j \in \mathcal{N}_i} \frac{1}{G} \sum_{g=1}^G \mathbf{w}_g(\mathbf{e}_{ij}) \odot \mathbf{\Theta}_g \mathbf{x}_j, \end{equation} where \begin{equation} \label{eq:GMMConvweight} \mathbf{w}_g(\mathbf{e}) = \exp \left( -\frac{1}{2} {\left( \mathbf{e} - \mathbf{\mu}_g \right)}^{\top} \Sigma_g^{-1} \left( \mathbf{e} - \mathbf{\mu}_g \right) \right).
\end{equation}
denotes a weight matrix with trainable mean vector $\mu_g$ and trainable
diagonal covariance matrix $\Sigma_g$. $\odot$ denotes the element-wise
product, $\mathbf{x}_i$ are the features of the network at node $i$, and
$\mathbf{x}^{\prime}_i$ are the outputs of the network at node $i$,
$\mathcal{N}_i$ denotes the neighborhood (connected nodes) of node $i$, $G$
denotes the number of Gaussian kernels used, and $\mathbf{\Theta}_g$ represents
the set of trainable weights convolved with the inputs for kernel $g$.
For our best model PID classification model, $G$ is set to 2.
The edge attribute $\mathbf{e}_{ij}$ is the normalized relative Cartesian coordinates between linked nodes:

\[\mathbf{e}_{ij} = \frac{1}{2\max_{(i,j)}(|\mathbf{p}_j - \mathbf{p}_i|)} (\mathbf{p}_j - \mathbf{p}_i) + 0.5\] where $\mathbf{p}_i$ denotes the position of node $i$ and $\max_{(i,j)}(|\mathbf{p}_j - \mathbf{p}_i|)$ is the maximum absolute difference across all dimensions and all edges.

This shared core is referenced in the following task-specific sections.

\subsection{PID Classification with GNNs} \label{subsec:PIDGRAPHCONV}
For PID classification (see Table~\ref{table:PID} for list of PIDs), the model utilizes the core GNN architecture described in Section~\ref{subsec:CoreGNNArchitecture}
with an output dimensionality of five per segment, corresponding to log-likelihoods for each PID class. The $k$-NN graph construction with $k=5$
is optimal. The output of a GNN layer preserves the multiplicity of each event.
This is unlike the CNN layer as described in Reference~\ref{subsec:CoreCNNArchitecture} which requires special care to retrieve outputs for the relevant segments
firing in a given event.

\begin{table}[htbp]
	\centering
	\caption{Architecture used for best performing GNN PID classifier.
		\textmd{Each GMMConv layer is followed by a batch normalization and then a ReLU
			layer.
			The first 130 input features are the concatenated 65 sample waveforms for left and right PMTs.
			The intermediate layers expand the feature size up to 435 then contract to the final 5 features representing class scores for each of the classes shown in table~\ref{table:PID}.
			The 5 nearest neighbors were used for constructing the adjacency matrix in this model.
			Cartesian distance weighting for the edges and a kernel size of 2 is used for every GMMConv layer.
		}}
	\smallskip
	\label{table:BestPIDGraphModel}
	\begin{tabular}{|l | l | l|}
		\hline
		\bf Layer Number & \bf Input Size & \bf Output Size \\
		\hline
		1                & 130            & 191             \\
		2                & 191            & 252             \\
		3                & 252            & 313             \\
		4                & 313            & 374             \\
		5                & 374            & 435             \\
		6                & 435            & 364             \\
		7                & 364            & 293             \\
		8                & 293            & 222             \\
		9                & 222            & 151             \\
		10               & 151            & 80              \\
		11               & 80             & 5               \\
		\hline
	\end{tabular}
\end{table}

\subsection{Position Reconstruction with GNNs} \label{subsec:graphz}
Predictions for $z$ using graph convolutional layers follow the same pattern as for PID predictions, employing the same core GNN architecture as described in Section~\ref{subsec:CoreGNNArchitecture} but with a scalar output per segment.
The approach to adjacency matrix selection is detailed, but performance does not exceed that of the CNN-based method.

It was found through several dozen trials of a hyperparameter search that the GNN does not perform as well as the CNN for $z$ predictions as measured by the cross-entropy loss over the validation dataset.
This is in contrast to PID prediction, which the GNN outperforms the equivalent CNN model.

In an attempt to match the performance of the CNN $z$ prediction described in Section~\ref{subsec:CNNPos}, an adjacency matrix was chosen that was identical to a square filter running over each of the segments.
This was done by choosing connections that are 1, 2 or 3 segments apart in x and y (corresponding to a filter size of 3 $\times$ 3, 5 $\times$ 5, and 7 $\times$ 7).
The pattern of changing between pointwise convolutions, which are graphs consisting of a single segment with self-loops, and 7-, 5- or 3-sized filters was copied.
Using this approach, roughly the same results as using the k-nearest neighbors approach were achieved, indicating that the choice of adjacency matrix was not the reason for the reduced performance compared to the CNN approach.
See Table~\ref{table:ZPerformanceComparison} for a full comparison of neural network architectures used for $z$ prediction.

\subsection{GNN Training and Optimization}
Training and optimization for GNNs follows the same procedure as described for CNNs in Section~\ref{subsec:CNN}.
The same datasets, loss functions, and optimization algorithms were used.
The only difference is in the specific hyperparameters that were varied for the optimization, which included the number of nearest neighbors for constructing the adjacency matrix and the type of GNN layer used.

\subsection{GNN Performance Summary}\label{sec:GraphPerformance}
The best performing GNN model for PID classification was shown in Table~\ref{table:BestPIDGraphModel}.
This model achieved an average false positive rate of 67.6\% for ionization selection at the optimal cut threshold, compared to 84.8\% for the conventional SE PSD based method and 72.3\% for the best CNN model.
A full comparison of the PID classification performance is shown in Figure~\ref{fig:PIDConfusion}.

For position reconstruction, the best performing GNN model achieved a mean absolute error of 147 mm on the first day of data used for the IBD analysis, compared to 133 mm for the best CNN model.
A detailed comparison of the performance of different architectures for position reconstruction is given in Section~\ref{sec:PerformanceComparison}.

\section{Other Machine Learning Techniques Explored} \label{sec:othertechs}
In addition to GNNs and CNNs, other architectures	were attempted.
The simplest type of neural network, a fully-connected neural network (FCNN) was applied to the waveforms prepared as described in Section~\ref{subsec:waveform-preparation}.
A fully connected neural network (FCNN) is one in which each input is connected to every output.
Each layer in this type of network can be described as a linear transformation, \begin{equation} \label{eq:linear} y = Ax + b.
\end{equation}
Here $y$ is the output of the layer, $x$ is the input, $A$ is a matrix of trainable weights and $b$ is a trainable constant.
These will be referred to as linear layers.
Between each linear layer, ReLU activation function was used, as with the other architectures presented in this work.
It was found through experimentation that the best models were those in which a number of linear layers are chained together, first expanding the input waveform into a size roughly twice that of the number of samples fed into the network.
This is followed by a number of layers that shrink the number of outputs to a single output which is the prediction.
FCNN were approximately as successful at predicting the $z$ position for events that have a multiplicity of one as GNNs and CNNs.
For higher multiplicity events, CNNs and GNNs that take in all data from neighboring segments to predict the single segment perform better than FCNN models.

Other, more complex architectures were attempted.
Because the waveform is a time series, it was initially thought that using architectures designed for processing time series data could help capture additional information to improve upon the naive approach used in the GNNs or CNNs, which was to simply feed each sample of the waveform as feature layers of the network.
To this end, recurrent neural networks (RNNs) and temporal convolutional networks~\cite{TCN} (TCNs) were used to process the waveforms before feeding them into a CNN architecture.
In the case of the TCN, the outcome was identical to no TCN being applied.
In the case of the RNN, the gradient descent algorithm was unable to converge and no useful predictions were able to be made.

Another architecture tried in the same vein was 3D convolutional networks where the 3rd axis used is the temporal axis of the waveform.
In this construction we use the two samples from both PMTs as the two features fed into the newtork at each time and spatial location.
The difficulty with this approach was the enormous amount of time it takes to perform 3D convolutional layers.
To decrease the computational complexity, we factorized the time and spatial dimensions by first performing pointwise convolutions in space and length N convolutions in time to reduce the size of the temporal dimension before performing spatial convolutions with a filter of length 1 in the time dimension.
This still resulted in a factor of 6 increase in CPU time per epoch relative to 2D convolutions so we did not perform detailed tests of the algorithm.

\section{Performance Comparison of Techniques for SEER}
\label{sec:PerformanceComparison}
Table~\ref{table:ZPerformanceComparison} shows the performance comparison between different NN architectures used for $z$ reconstruction.
%The error shown is for the best network found for the given architecture after a hyperparameter search. Typical hyperparameter searches explored at least 20 different trials, although more trials were performed for those that showed more promise such as the CNNs and GCNs. Note the nearest neighbor average technique is not a NN but a simple averaging of locations from hits in neighboring DE segments. 
It is interesting to note that the GNN does not perform as well as the CNN for this problem, despite outperforming the CNN in particle identification.
This could likely be improved by using a graph layer tailor-made specifically for this problem rather than an off-the-shelf algorithm.
It is also interesting to note that when using extracted features such as pulse area and timing, the network performs significantly worse.
This performance improves when adding more features, although it is still quite far behind the full waveform models.

The single waveform model only uses information from the segment of the energy deposition of interest.
The fact that this model performs better than the neighboring average method indicates that the pulse itself contains information about the position which is not being captured in the extracted features.
Given the gap in performance between the extracted features models and the full waveform models, more work is needed to determine if there are parameters that could be extracted to better predict the SE $z$ position.

\begin{table}[htbp]
	\centering
	\caption{Performance comparison between different NN architectures used
		for $z$ reconstruction. \textmd{The error shown is for the best network found
			for the given architecture after a hyperparameter search.
			Typical hyperparameter searches explored at least 20 different trials, although more trials were performed for those that showed more promise such as the CNNs and GNNs.
			Note the nearest neighbor average technique is not a NN but a simple averaging of locations from hits in neighboring DE segments.
		}}
	\smallskip
	\label{table:ZPerformanceComparison}
	\begin{tabular}{|l | l | l |}
		\hline
		\bf Architecture             & \bf Mean Absolute Error [mm] &
		\bf Description                                               \\
		\hline
		CNN with Waveform            & 133                          &
		\ref{subsec:CNNPos}                                           \\
		GNN with Waveform            & 147                          &
		\ref{subsec:graphz}                                           \\
		CNN with Extracted Features+ & 212                          &
		\ref{subsec:extracted-quant}                                  \\
		Single Waveform FCNN         & 218                          &
		\ref{sec:othertechs}                                          \\
		CNN with Extracted Features  & 238                          &
		\ref{subsec:extracted-quant}                                  \\
		Nearest neighbors average    & 306                          &
		\\
		\hline
	\end{tabular}
\end{table}

\begin{figure}[htbp] \centering
	\includegraphics[width=1.0\linewidth]{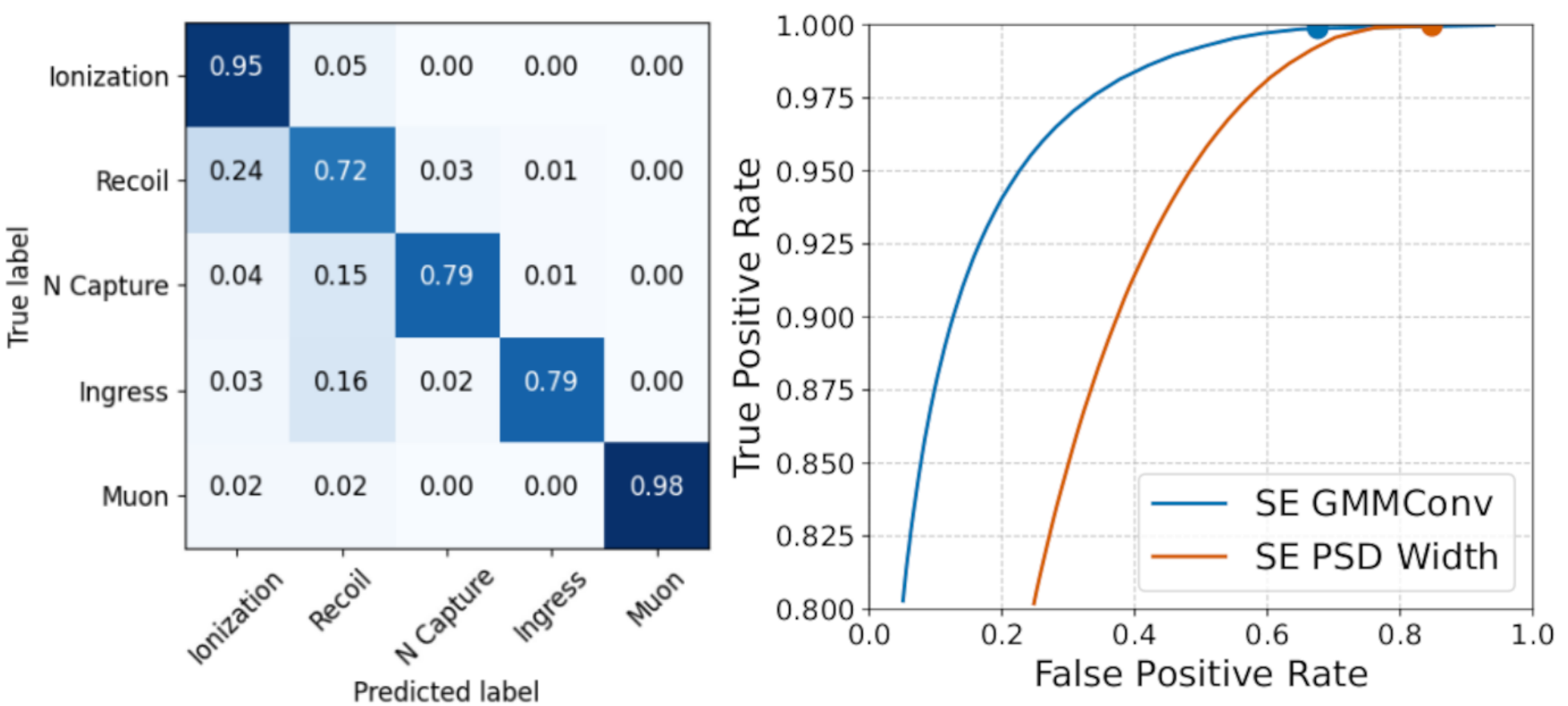}

	\caption{The confusion matrix for the GMMConv classifier (see section ~\ref{subsec:PIDGRAPHCONV}) is
		shown on the left.
		Most problematic for the IBD analysis are recoil events which are mislabeled ionizations.
		The Receiver-operator-characteristic curve for ionization selection is shown on the right.
		Curves are shown for the GMMConv classifier and the non-ML based classifier using the SE PSD distribution as described in Reference~\ref{sec:SEER}.
	}
	\label{fig:PIDConfusion}
\end{figure}

Figure~\ref{fig:PIDConfusion} shows the performance comparison for PID classification between the best performing GNN model (GMMConv) and the conventional SE PSD based method.
On the left is a confusion matrix showing the percentage of single-ended PIDs that were predicted for each category.
The horizontal axis is the predicted label and the vertical axis is the true label.
The numbers are divided by the total number of true labels for each row.
The predicted label is the one with the largest class score output by the model.
From this table we can see that 24\% of recoils are misidentified as ionizations, constituting the largest source of background for the event selection.
Conversely, 5\% of ionizations are misidentified as recoils.

On the right is a plot showing the receiver-operator-characteristic (ROC) curve for ionization selection.
Here a binary choice problem is solved in two ways, one using ML and the other using the traditional PSD analysis.
For the ML method, a threshold for the GMMConv classifier ionization score is chosen above which we consider the deposition an ionization.
This threshold is varied in order to produce the depicted curve of true positive rate (TPR - percentage of ionizations correctly labeled) vs false positive rate (FPR - percentage of non-ionizations labeled as ionization).
Also depicted is the ionization classification using the SE PSD distribution, varying the number of sigma above the mean SE PSD as described in Section~\ref{sec:SEER} to produce the curve.
The marked points show the TPR and FPR for the threshold that optimizes the effective statistics of the dataset.
This will be described in more detail in the following section.
For the PSD based cut, this is at TPR = 99.97\%, and FPR = 84.82\%.
For the GMMConv classifier, this is at TPR = 99.88\% and FPR = 67.59\%.

\section{Impact on Inverse Beta Decay Selection in PROSPECT}
\label{sec:ML_IBD_selection}
This section will describe the usage of the ML-based SEER for enhancing IBD background discrimination.
The nominal SEER cuts based on Gaussian fits to the SE PSD band at different energies and the SE visible energy (no position correction) is described in detail in Section~\ref{sec:SEER}.
A comparison between the nominal SEER cuts and the ML-based method will be shown.
A discussion on further possible improvements given more detailed waveform simulation development is given at the end of the section.

\subsection{Method} \label{sec:ML_IBD_section_method}

The ML approach follows as described in Section~\ref{sec:ML-SEER}.
Position reconstruction using a CNN is used to estimate the energy which allows a tighter bound on the largest allowed energy hit in a SE segment.
Ionization identification using a GNN allows us to replace the SE PSD fit-based cut with a class score threshold.
Events with any SE segments containing a ionization class score below the threshold are rejected.
Additionally, a veto window is constructed identical to the ``SE Recoil veto'' as described in Section~\ref{sec:SEER} for events with an ionization class score below this threshold.
The model used for energy reconstruction is shown in Table~\ref{table:BestZModelDiagram} and the event classification model is shown in Table~\ref{table:BestPIDGraphModel}.

In order to optimize the energy threshold and class score threshold, first the nominal SEER cuts~\cite{PROSPECT:2022wlf} as shown in Table~\ref{tab:IBD_cuts} are used and the energy cut or the PSD cut are individually optimized by replacing the non-ML SEER cut with the ML-based cut one at a time.
The metric used for maximization is the effective statistics $N_{\mathrm{eff}}$ which is defined as

\begin{equation} \label{eq:effective_stats} N_{\mathrm{eff}} = \sum_i n_i^2 / \sigma_i^2, \end{equation} where $n_i$ is the background subtracted IBD counts in energy bin $i$ and $\sigma_i$ is the statistical uncertainty based on counting statistics for the accidentals, and the reactor-on and -off statistics.
Note that backgrounds are estimated by applying identical cuts to the reactor-off data.
There is a trade-off between increasing the total reactor-on statistics which increases the effective statistics but also increases the reactor-off total statistics that in turn lowers the effective statistics due to the additional uncertainty in the background induced.
Also, accidental backgrounds are subtracted by estimating their rate using shifted coincidence windows; stricter cuts reduce both the accidental rate and its associated uncertainty, thereby increasing the effective statistics.

Once the value of the energy and classifier thresholds are chosen based on maximization of the effective statistics for 20\% of the dataset sampled evenly across all 5 reactor cycles for which data was taken, the remaining cuts shown in Table~\ref{tab:IBD_cuts} were varied one at a time to maximize the effective statistics while the ML-based cuts are set in place.
Finally, once the new values of the non-ML cuts are found, the ML cuts are then re-varied for the final maximum using the full dataset shown also in Table~\ref{tab:IBD_cuts}.

\begin{table}[htbp]
	\centering
	\caption{Table of cuts used for ML and non-ML based SEER analysis.
		`Delayed PSD' is the \el{Li}{6} neutron capture sigma width for
		the delayed signal (which must be a single segment cluster), while `Delayed E'
		is the sigma width of its energy.
		`Prompt PSD' is the sigma width of the electron-like band,
		`Muon E' is the energy above which a cluster is vetoed with a window shown by
		`Muon Veto'. `Distance' is the maximum allowable distance in $z$ between the
		prompt and delayed for delayed occurring in the (same, adjacent) segment as the
		prompt.
		`SE PSD' is the sigma width of the ionization-like SE PSD, `ML
		Score' is the GMMConv ionization classification score for SE segments, `SE E' is the
		SE energy upper limit for SE segment hits, and `ML SE E' is the SE CNN energy
		prediction for those hits.}
	\smallskip
	\label{tab:IBD_cuts}
	\centering
	\begin{tabular}{|l|l|l|}
		\hline
		Name         & Nominal Value~\cite{PROSPECT:2022wlf} & ML Value
		\\ \hline
		Delayed PSD  & 2.2 $\sigma$                          & 2.2
		$\sigma$                                                        \\
		Delayed E    & 2.0 $\sigma$                          & 2.0
		$\sigma$                                                        \\
		Prompt PSD   & 2.0 $\sigma$                          & 2.1
		$\sigma$                                                        \\
		Muon E       & 15.0 $\mevee$                         & 18.0
		$\mevee$                                                        \\
		distance     & (140, 60) mm                          & (140,
		70) mm                                                          \\
		Fiducial $z$ & 968 mm                                & 984 mm
		\\
		Pileup Veto  & 800 ns                                & 800 ns
		\\
		Recoil Veto  & [0, 200] us                           & [0, 190]
		us                                                              \\
		nCap Veto    & [-300, 300] us                        & [-270,
		270] us                                                         \\
		Muon Veto    & [0, 200] us                           & [0, 180]
		us                                                              \\
		SE PSD       & 3.5 $\sigma$                          & ---
		\\
		ML Score     & ---                                   & 0.035
		\\
		SE E         & 0.8 $\mevee$                          & ---
		\\
		ML SE E      & ---                                   & 0.6
		$\mevee$                                                        \\
		\hline
	\end{tabular}
\end{table}

\subsection{Results}
After the first round of optimization using the nominal values for all non-ML cuts and using the CNN SE energy estimator, it was found that the SE energy threshold could be moved from 0.8 SE \mevee~to 0.6 ML \mevee.
This resulted in a small 0.5\% increase in the effective statistics of the dataset.

For the GMMConv classifier, initial testing was performed on a dataset with uniform background sampling, which contained a higher proportion of ionizations than typically seen in IBD candidate events (which require delayed neutron capture).
On this test dataset, a class score threshold of 0.21 was found to optimize the F1 score, a metric that balances precision and recall.
However, when applying the GMMConv classifier to actual IBD selection, a much lower threshold of 0.03 was found to be optimal, resulting in a 1.5\% gain in effective statistics.
This difference in optimal thresholds is consistent with the different ionization proportions between the test and application datasets.

Introducing both of these ML cuts with the nominal non-ML cut values, a 2\% increase in effective statistics results along with a 8\% increase in the signal-to-accidental background ratio and a 4\% increase in signal-to-correlated background ratio.

This increase in signal-to-background allows us to relax the non-ML cuts to less restrictive values which allows more of both signal and background.
Because the ML SEER cuts are more efficient than the non-ML SEER cuts, this increase in both signal and background brings about larger effective statistics.
The final re-optimized cut values are shown in the last column of Table~\ref{tab:IBD_cuts}.
Ultimately, the introduction of the ML SEER cuts results in a 4\% increase in effective statistics, albeit with a 2\% (4\%) reduction in signal-to-correlated (accidental) background ratio.
The results are summarized in Table~\ref{tab:ml_ibd_sel_results}.

This is a small performance increase given the much improved analysis offered by the machine learning.
This is due to the fact that the SE analysis is only aimed at rejecting backgrounds.
While the ML techniques yield significantly more powerful background rejection as shown in figure~\ref{fig:PIDConfusion}, the optimal effective statistics were found at a very low discrimination threshold due to the much larger amount of real signals compared to backgrounds.
This is demonstrated by the fact that the optimal ionization score was a very low 0.035 (1.0 roughly equating to 100\% confidence of the event being that class) which yielded a high FPR of 68\% in our test dataset.
Higher thresholds of the ionization score excluded too much of the real signal which greatly outweighed the backgrounds, a testament to the high selectivity of the prompt-delayed signature used in the experiment.

\begin{table}[htbp]
	\centering
	\caption{Results for IBD selection using only double-ended detectors
		(no SE), nominal SEER values, nominal values with ML SEER subbed in, and
		reoptimized values with ML SEER.
		A 3.3\% increase in effective statistics is observed with the introduction of ML SEER.
	}
	\smallskip
	\label{tab:ml_ibd_sel_results}
	\begin{tabular}{|l|l|l|l|l|}
		\hline
		                               & No SE      & Nominal & Nominal + ML
		                               & ML Optimal                          \\ \hline
		IBD Stats / Day                & 529.0      & 497.0   & 487.4
		                               & 518.7                               \\
		Signal : Correlated Background & 1.37       & 3.33    & 3.50
		                               & 3.27                                \\
		Signal : Accidental Background & 1.78       & 3.98    & 4.29
		                               & 3.82                                \\
		Effective Stats / Day          & 160.0      & 244.9   & 248.0
		                               & 253.1                               \\
		\hline
	\end{tabular}

\end{table}

\subsection{Discussion}

Modest improvements in the event selection were obtained using machine learning based SEER.
It was found that using the full information in the waveform was crucial for extracting as much information as possible from the SE segment data, which was the enabling factor for these improvements.

%One interesting finding in this study is that both graph and convolutional neural networks exhibit similar performance for tasks of interest related to event reconstruction.
%It is not entirely clear why the CNN outperforms the GNNs on $z$ reconstruction while the GNN outperforms the CNN on segment event classification. It was also found that different types of graph network implementations worked to varying degrees for different problems. The GMMConv that was chosen as the best classifier, for example, did not perform best at $z$ reconstruction. Thus, we find that it is valuable to experiment with different implementations of ML algorithms to find which ones best suit the problem at hand.

Future work could expand on this result by using ML-based classification on cluster-level data to determine if it is positron-like.
Because we do not have a curated dataset of positron events, the classifier would have to be trained on simulated events.
This would require more detailed Monte Carlo-based pulse simulation that accurately captures the underlying dynamics of the scintillation light production, propagation, and detection.
Such computational models would need validation against the existing detector dataset, using pulse characteristics including height, width, PSD, rise time and fall time as a function of position, particle type, and energy.
In this work, we circumvented the need for synthetic pulse generation by implementing a fully data-driven approach, training directly on experimental data and using the dual-ended reconstructed calibrated values as the target.

A realistic pulse simulation would also have the benefit of aiding in the calculation of the ML-based event selection efficiency and the covariance matrix of the energy estimation.
These are both needed to properly estimate the uncertainties in the IBD energy distribution and extracted fits for sterile oscillation parameters.

It is likely that similar architectures could be applied to other particle detector experiments that utilize highly segmented geometries with light readouts on the detector ends.
By using the full waveforms from the light readouts once can utilize the maximum amount of information available to best determine the dynamics of the interacting particles and thereby increase classification power and improve spatial and energy resolution.

\section{Conclusion}
Convolutional neural networks and graph convolutional networks were applied to the PROSPECT data for the purpose of recovering information from SE segments.
It was found that using the full waveform information from the digitized PMT signals was required to provide the best performance for SE reconstruction.
CNNs were found to excel at reconstructing the position of the event in SE segments while GNNs performed best at classifying energy depositions in SE segments.
With a position reconstruction one can better estimate the amount of energy deposited in a segment and use this to more powerfully reject IBD backgrounds.
Similarly, the classification reconstruction can also help determine heavier recoiling particles in SE segments which allows further background rejection.
It was shown that the GNN outperformed the tail fraction method of PSD classification for SE segments.
This resulted in a larger overall signal to background, which allowed us to relax veto window lengths so as to maximize the effective statistics of the dataset using ML.
A 3.3\% improvement in effective statistics was achieved over the traditional tail fraction method of particle identification.
The demonstrated effectiveness of ML techniques to improve event reconstruction on PROSPECT opens up promising avenues for its application for similar segmented particle detectors, such as PROSPECT-II.